\documentclass{evn2002}
\begin{document}
   \title{EVN and MERLIN confirmation of the \object{LS~5039} jets}

   \author{J.~M. Paredes\inst{1}
          \and M. Rib\'o\inst{1}
          \and E. Ros\inst{2}
          \and J. Mart\'{\i}\inst{3}
          \and M. Massi\inst{2}
}

   \institute{Departament d'Astronomia i Meteorologia, Universitat de Barcelona, Av. Diagonal 647, 08028 Barcelona, Spain
\and Max Planck Institut f\"ur Radioastronomie, Auf dem H\"ugel 69, 53121 Bonn, Germany
\and Departamento de F\'{\i}sica, Escuela Polit\'ecnica Superior, Universidad de Ja\'en, Virgen de la Cabeza 2, 23071 Ja\'en, Spain
}

   \abstract{
The microquasar nature of \object{LS~5039} was revealed 
by May 1999 VLBA+VLA observations showing a two-sided jet 
at milliarcsecond scales. Here we present
follow-up interferometric observations carried out with the EVN and MERLIN at
5~GHz in March 2000. The obtained maps with both the EVN and MERLIN show a
two-sided jet with a similar position angle to the previous VLBA+VLA map. The
total length of the jet arms is $\sim60$~mas 
in the EVN map and $\sim300$~mas in
the MERLIN map. A brightness and length asymmetry of the jets, compatible with
the earlier observations, is also present in the maps. Overall, these
observations confirm the existence of a two-sided jet structure in
\object{LS~5039} and seem to indicate their persistent nature.
   }

   \maketitle
%

\section{Introduction}

\object{LS~5039} is a high mass X-ray binary system, with 
an optical magnitude $V=11.2$
and spectral type O6.5V(f), located at a distance of $\sim2.9$~kpc and close
to the galactic plane ($l=16.88\degr$, $b=-1.29\degr$). VLA observations
carried out by Mart\'{\i} et~al. (\cite{marti98}) found that the source
was also a non-thermal radio emitter with moderate variability. Paredes et~al.
(\cite{paredes00}) discovered that the system displays relativistic radio jets,
revealing the microquasar nature of \object{LS~5039}, and proposed an
association with the high energy $\gamma$-ray source \object{3EG~J1824$-$1514}.
The population of microquasars is still a very reduced one, with the best
representative examples being \object{SS~433}, \object{GRS~1915+105},
\object{Cyg~X-3} and \object{GRO~J1655$-$40} (Mirabel \& Rodr\'{\i}guez
\cite{mirabel99}). McSwain et~al. (\cite{mcswain01}) have recently obtained the
radial velocity curve of the system, determining a period of $P\simeq4.1$~days
and a high eccentricity of $e\simeq0.4$. Recently, Rib\'o et~al.
(\cite{ribo02}) have found that \object{LS~5039} is a runaway X-ray binary.

\section{Observations and data reduction}

We observed \object{LS~5039} simultaneously with MERLIN and the EVN on March
1st 2000 (3:20--7:10~UT) at 5~GHz. Single dish flux density measurements were
carried out with the MPIfR 100~m antenna in Effelsberg, Germany.


The EVN observations were performed with EB, JB, CM, WB, MC, NT, and TR,
recording in MkIV mode with 2~bit sampling at 256~Mbps at left hand circular
polarization, allowing a bandwidth of 64~MHz. The data were processed at the
MkIV correlator at JIVE with an integration time of 4~s. Interferometer fringes
for \object{LS~5039} were detected in all baselines. A later fringe fitting of
the residual delays and fringe rates was performed within {\sc aips} for
\object{LS~5039}. We averaged in frequency the data and exported them to be
imaged and self-calibrated into {\sc difmap}. The final imaging was carried out
on those data after editing and averaging of the visibilities in 32~s blocks.

The EVN and MERLIN arrays have one common baseline, from JB to CM. That allows
to combine both data sets and find some redundancy in the data to map them
together, reaching $(u,v)$ resolution ranges from 0.015~M$\lambda$ (MK2-Tabley)
to 26~M$\lambda$ (JB-NT) at 5~GHz.

\begin{figure*}[htpb]
\vspace{315pt}
\includegraphics{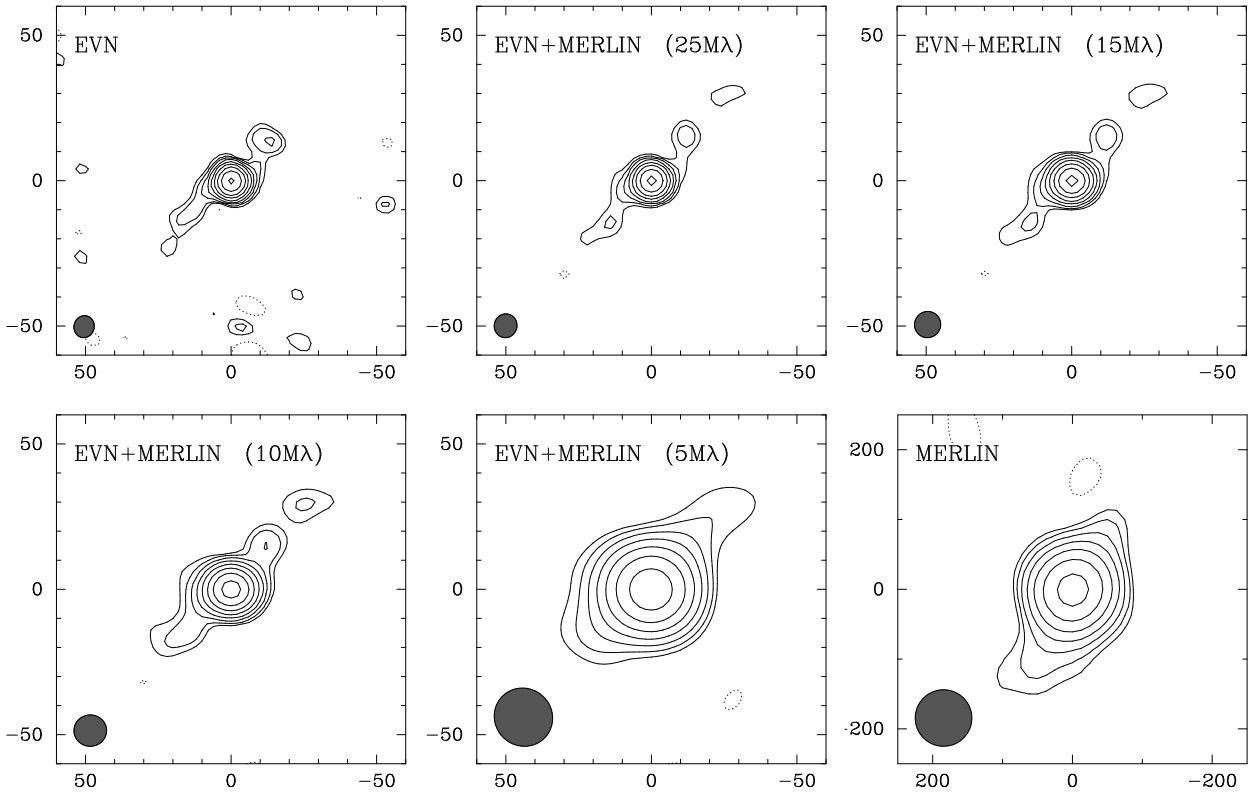}
\caption{Self-calibrated maps of \object{LS~5039} at 5 GHz obtained on March 1, 2000, from higher to lower resolutions. 
Contours are those listed as $S_\mathrm{min}$ in Table~\ref{table:param} and scaled with $\sqrt3$.
Note that the map scale is different at the bottom right panel 
(MERLIN data).  }
\label{fig:allmaps}
\end{figure*}

\section{Results and discussion}

The flux density monitoring with the Effelsberg antenna was not reliable 
due to confusion to nearby sources.
However, inspection of the shortest baselines of MERLIN
reveals a constant flux density during the full observation. We present all the
resulting maps in Fig.~\ref{fig:allmaps}, and the parameters of each one in
Table~\ref{table:param}. These maps clearly show that \object{LS~5039} is a
source of bipolar jets emanating from a central core. There is some asymmetry
in the jets, both in flux density and separation from the core, that may
involve relativistic beaming. These results confirm the existence of bipolar
radio jets in \object{LS~5039} obtained in previous VLBA+VLA observations by
Paredes et~al. (\cite{paredes00}). We must note that this source
apparently does not show
strong outbursts or, at least, have never been detected 
in the eleven-month monitoring carried out by the Green Bank 
Interferometer. This suggests that the jets could be steady, 
as seem to indicate the VLBI maps obtained up
to now. The total size of the jets in the EVN map is $\sim$180~AU and 
$\sim$900~AU in the MERLIN map. Hence, the jets extend to larger distances
than those typically imaged with the VLBA+VLA observations, of $\sim$18~AU. All of them
have similar position angles, being at a P.A.\ of $\sim125\degr$ at VLBA scales, P.A.
$\sim140\degr$ at EVN scales and P.A.\ $\sim150\degr$ at MERLIN scales,
suggesting a bending of the jets with increasing distance from the core and/or
precession. From the EVN map, we obtain for the approaching component
(southeast) a flux density of 1.8 mJy and a total distance from the core of 34
mas. For the receding component (northwest) the values are 1.5 mJy and 24 mas. 
Using these values, and assuming that the length asymmetry is due to Doppler
boosting, we estimate lower limit of the jet speed of $\beta>0.17\pm0.05$ and an
upper limit for the viewing angle of the jet of
$\theta<80\degr\pm3\degr$. These values are similar to those found in the
VLBA+VLA map. A detailed analysis can be found in Paredes et~al. (in
preparation).

\begin{table}[htbp]
\caption[]{Image parameters in Fig.~\ref{fig:allmaps}}
\begin{center}
\begin{tabular}{@{}l@{~~}c@{~~}c@{~~}c@{~~}c@{}}
\hline \hline \noalign{\smallskip}
Array                       & beam size             & P.A.      & $S_{\rm peak}$ & $S_{\rm min}$ \\ 
                            & [mas]$\times$[mas]    & [$\degr$] & [mJy]        & [mJy] \\
\noalign{\smallskip} \hline \noalign{\smallskip}
EVN                         & 7.60$\,\times\,$6.96              & $-$14        & 26.6 & 0.3 \\
EVN+MERLIN (25\,M$\lambda$) & 8.21$\,\times\,$7.80  & $-$14     & 26.8         & 0.5 \\
EVN+MERLIN (15\,M$\lambda$) & 9.08$\,\times\,$8.93  & $-$39     & 26.7         & 0.5 \\
EVN+MERLIN (10\,M$\lambda$) & 11.2$\,\times\,$10.8  & $-$87     & 26.5         & \,~0.45 \\
EVN+MERLIN (5\,M$\lambda$)  & 20.4$\,\times\,$19.7  &  ~~47     & 26.7         & 0.7 \\
MERLIN          & ~\,81.0$\,\times\,$81.0$^{\rm a}$ & \,~~~0    & 33.4         & 1.0 \\
\noalign{\smallskip} \hline
\end{tabular}
\end{center}
\begin{list}{}{
}
\item[$^{\rm a}$] This is a circular beam, equivalent to the interferometric synthesized beam of 142$\times$46~mas (P.A.~$-$47\degr).
\end{list}
\label{table:param}
\end{table}

\begin{acknowledgements}

The European VLBI Network is a joint facility of European, Chinese 
and other
radio astronomy institutes funded by their national research councils.
This research was supported by the European Commission's TMR and IHP Programme
``Access to Large-scale Facilities", under contract No.\ ERBFMGECT950012 and HPRI-CT-1999-00045, respectively.
We acknowledge the support of the European Community - Access to Research
Infrastructure action of the Improving Human Potential Programme.
J.~M.~P., M.~R., and J.~M. acknowledge partial support by DGI of the Ministerio de Ciencia y Tecnolog\'{\i}a (Spain) under grant AYA2001-3092, as well as partial support by the European Regional Development Fund (ERDF/FEDER).
M.~R. is supported by a fellowship from CIRIT (Generalitat de Catalunya, ref. 1999~FI~00199).
J.~M. is partially supported by the Junta de Andaluc\'{\i}a and by an Henri Chr\'etien International Research Grant (AAS).

\end{acknowledgements}


\begin{thebibliography}{}

\bibitem[1998]{marti98}
Mart\'{\i}, J., Paredes, J.~M., \& Rib\'o, M.
1998, A\&A, 338, L71

\bibitem[2001]{mcswain01}
McSwain, M.~V., Gies, D.~R., Riddle, R.~L., Wang, Z., \& Wingert, D.~W.
2001, ApJ, 558, L43

\bibitem[1999]{mirabel99}
Mirabel, I.~F., \& Rodr\'{\i}guez, L.~F.
1999, ARA\&A, 37, 409

\bibitem[2000]{paredes00}
Paredes, J.~M., Mart\'{\i}, J., Rib\'o, M., \& Massi, M.
2000, Science, 288, 2340

\bibitem[2002]{ribo02}
Rib\'o, M., Paredes, J.~M., Romero, G.~E., et~al.
2002, A\&A, 384, 954

\end{thebibliography}
\end{document}